# Spectral Analysis and Liouville–Green Asymptotics for a Radial Sturm–Liouville Operator with Variable Coefficients

**Victor S. Gerasimchuk[1], Bohdan A. Yevdokymenko[1], Igor V. Gerasimchuk[2,3]**

[1] Faculty of Physics and Mathematics, National Technical University of Ukraine “Igor Sikorsky Kyiv Polytechnic Institute”, Beresteisky Ave. 37, 03056 Kyiv, Ukraine
E-mail: viktor.gera@gmail.com

[2] Department of Physics of Meso- and Nanocrystal Magnetic Structures, V.G. Baryakhtar Institute of Magnetism of the NAS of Ukraine, Vernadsky Blvd. 36b, 03142 Kyiv, Ukraine

[3] Institute of Physics and Technology, National Technical University of Ukraine “Igor Sikorsky Kyiv Polytechnic Institute”, Beresteisky Ave. 37, 03056 Kyiv, Ukraine

**Abstract:** We investigate a radially symmetric initial-boundary value problem whose separation of variables leads to a regular self-adjoint Sturm–Liouville problem with explicitly determined coefficients and a positive weight function. The resulting Sturm–Liouville operator is defined by a special geometry-induced coefficient pair that gives rise to a nonstandard weighted spectral structure. The associated radial operator possesses a discrete real spectrum and a complete orthonormal system of eigenfunctions in the corresponding weighted Hilbert space, yielding an exact spectral representation of the evolution problem. An exact transcendental spectral equation governing the eigenvalues is derived. To analyze the high-frequency regime, the Liouville–Green transformation is applied directly to the radial Sturm–Liouville equation. This yields explicit asymptotic formulas for the eigenvalues and eigenfunctions together with corresponding Liouville–Green quasimodes. Their weighted orthogonality properties and asymptotic completeness are established within the spectral framework of the exact operator. The resulting asymptotic spectral data are used to construct approximate solutions of the original boundary-value problem and to obtain error estimates for the spectral reconstruction. Numerical computations of the exact Sturm–Liouville spectrum show excellent agreement with the Liouville–Green predictions, thereby validating the asymptotic quantization formula and confirming the accuracy of the proposed spectral approximation.

## 1. Introduction

Second-order differential operators with radial coefficients naturally arise after separation of variables in wave propagation, vibration theory, and spectral problems possessing cylindrical or spherical symmetry. In the classical constant-coefficient setting, such reductions lead to the standard radial Laplace operator and the associated Bessel-type spectral theory. More general radial geometries generate Sturm–Liouville operators with variable coefficients whose spectral properties are determined by the underlying coefficient structure.

The classical theory of regular Sturm–Liouville operators on finite intervals provides a rigorous framework for the analysis of such problems [1–5]. Under standard assumptions on the coefficients and boundary conditions, the spectrum is real and discrete, the eigenfunctions form a complete orthonormal basis in an appropriate weighted Hilbert space, and the corresponding evolution equations admit exact spectral decompositions. These properties form the mathematical foundation of many spectral methods in mathematical physics.

In many applications, however, the coefficients of the radial operator are sufficiently complicated that explicit closed-form spectral solutions are unavailable. In such situations, asymptotic techniques become essential for describing the high-frequency spectral regime. Among these, the Liouville–Green (WKB) method provides a powerful tool for constructing asymptotic approximations of eigenvalues and eigenfunctions while preserving the underlying Sturm–Liouville structure [6–9].

The present paper studies a radial Sturm–Liouville operator with variable coefficients arising from a geometry-induced radial structure motivated by models of confined oscillatory media and topological spin textures (see, e.g., [10,11]). After separation of variables, the problem reduces to a regular Sturm–Liouville equation with explicitly determined coefficients and a strictly positive weight function. Although the resulting operator is regular on the interval under consideration, the specific Sturm–Liouville coefficient pair generated by the radial geometry possesses several distinctive features. In particular, it induces a nontrivial Liouville coordinate, an associated logarithmic radial metric, and a weighted spectral structure that differs substantially from the classical Bessel-type setting.

The radial equation has the form

$$\frac{d^2R}{dr^2}-\frac{1}{r}\frac{dR}{dr}+\lambda^2\left(\frac{r^2}{r_0^2}-1\right)R=0$$

posed on the finite interval

$$r\in[r_1,r_2],\qquad r_1>r_0>0,$$

subject to homogeneous Dirichlet boundary conditions. Since the singular point of the coefficients lies outside the interval under consideration, the corresponding Sturm–Liouville problem remains regular throughout the domain.

The main objective of the paper is to provide a complete spectral analysis of this variable-coefficient radial Sturm–Liouville operator and to construct explicit Liouville–Green asymptotic approximations for its spectral data. We first establish the self-adjoint Sturm–Liouville framework associated with the problem, identify the corresponding weighted Hilbert space, and derive an exact spectral representation of the original initial-boundary value problem. An exact transcendental equation governing the eigenvalues is obtained from the boundary conditions.

To analyze the high-frequency regime, we apply the Liouville–Green transformation directly to the radial Sturm–Liouville equation. This yields explicit asymptotic formulas for the eigenvalues and eigenfunctions together with corresponding Liouville–Green quasimodes. We show that these asymptotic modes remain compatible with the weighted spectral structure of the exact operator and asymptotically reproduce the corresponding orthogonality relations. The completeness of the resulting quasimode system is established in Appendix A, while the uniform convergence of the associated modal expansions under suitable smoothness assumptions is proved in Appendix B.

The resulting asymptotic spectral data are then incorporated into the reconstruction of solutions of the original boundary-value problem. Error estimates are obtained for the asymptotic reconstruction, while oscillatory integral estimates controlling off-diagonal spectral interactions are established in Appendix C. Numerical computations of the exact Sturm–Liouville spectrum are finally used to validate the asymptotic quantization formula and to provide an independent assessment of the accuracy of the Liouville–Green approximation.

The principal results of the paper may be summarized as follows:

I. Reduction of the radial problem to a regular self-adjoint Sturm–Liouville operator with a strictly positive weight.

II. Construction of the exact spectral representation and characterization of the associated spectrum.

III. Derivation of explicit Liouville–Green asymptotic formulas for the spectral data, including eigenvalues and eigenfunctions.

IV. Development of an asymptotic spectral reconstruction of solutions based on Liouville–Green quasimodes.

V. Numerical validation of the asymptotic spectral approximations through comparison with the exact Sturm–Liouville spectrum.

The paper is organized as follows. Section 2 presents the formulation of the initial-boundary value problem and the underlying assumptions. Section 3 establishes the associated Sturm–Liouville framework and proves the self-adjointness of the radial operator. Section 4 develops the exact spectral representation of the evolution problem. Section 5 derives the Liouville–Green asymptotic formulas for the eigenvalues and eigenfunctions. Section 6 investigates the asymptotic spectral properties of the Liouville–Green quasimodes. Section 7 develops the asymptotic reconstruction of solutions and derives corresponding error estimates. Section 8 presents a numerical validation of the asymptotic spectral formulas. Technical results concerning completeness, convergence, and oscillatory integral estimates are collected in Appendices A–C.

## 2. Formulation of the Initial-Boundary Value Problem

We consider the radial initial-boundary value problem on the finite interval $r \in [r_1, r_2]$, $r_1 > r_0 > 0$. The condition $r_1 > r_0$ guarantees that the singular point of the coefficient lies outside the domain under consideration. Consequently, all coefficients of the reduced radial problem remain regular on the closed interval $[r_1, r_2]$.

The governing equation has the form

$$\frac{1}{c^2}\frac{\partial^2 u}{\partial t^2} = \frac{1}{\dfrac{r^2}{r_0^2} - 1}\left(\frac{\partial^2 u}{\partial r^2} - \frac{1}{r}\frac{\partial u}{\partial r}\right), \qquad r_1 < r < r_2, \quad t > 0. \tag{2.1}$$

We impose homogeneous Dirichlet boundary conditions

$$u(r_1, t) = 0, \quad u(r_2, t) = 0, \qquad t > 0, \tag{2.2}$$

together with the initial conditions

$$u(r, 0) = \varphi(r), \quad \left.\frac{\partial u(r,t)}{\partial t}\right|_{t=0} = \psi(r), \qquad r_1 < r < r_2. \tag{2.3}$$

The homogeneous boundary conditions are chosen in order to emphasize the spectral structure generated by the radial operator. Under these conditions, the spatial problem admits a natural formulation in terms of a regular Sturm–Liouville operator with separated boundary conditions.

The coefficient

$$\frac{1}{\dfrac{r^2}{r_0^2} - 1}$$

remains bounded on $[r_1, r_2]$ due to the assumption $r_1 > r_0$. Consequently, equation (2.1) defines a regular hyperbolic initial-boundary value problem on the finite interval.

A central role in the subsequent analysis is played by the associated weighted Sturm–Liouville structure obtained after separation of variables. In particular, the radial geometry generates a nontrivial weight function and induces the corresponding weighted Hilbert space in which the radial operator becomes self-adjoint.

We now state the standing assumptions used throughout the paper.

**Assumption 2.1 (Regular Sturm–Liouville structure).** The spatial differential expression associated with equation (2.1) has real-valued $C^1$-coefficients on the closed finite interval $[r_1, r_2]$. The induced Sturm–Liouville weight function $\rho(r)$, defined later in equation (3.4), is strictly positive and bounded away from zero on $[r_1, r_2]$. The boundary conditions (2.2) are regular and separated.

Under the above assumptions, the associated Sturm–Liouville realization is regular and self-adjoint in the corresponding weighted Hilbert space (see, e.g., [1,2,4]).

The formulation above provides the exact operator framework required for the spectral analysis developed in the subsequent sections. In particular, the positivity of the induced weight function will play an essential role in the construction of the exact spectral representation and in the derivation of the large-mode semiclassical asymptotics.

## 3. Reduction to a Sturm–Liouville Spectral Problem

Applying separation of variables to the problem (2.1)–(2.3) in the form $u(r,t) = R(r)T(t)$, we obtain

$$\frac{1}{c^2 T(t)} \frac{d^2 T}{dt^2} = \frac{1}{\frac{r^2}{r_0^2} - 1} \frac{1}{R(r)} \left( \frac{d^2 R}{dr^2} - \frac{1}{r} \frac{dR}{dr} \right).$$

Since the left-hand side depends only on $t$, whereas the right-hand side depends only on $r$, both sides must be equal to the same separation constant. Introducing the spectral parameter $-\lambda^2$, we arrive at the temporal equation

$$\frac{d^2 T}{dt^2} + c^2 \lambda^2 T = 0, \tag{3.1}$$

and the radial equation

$$\frac{d^2 R}{dr^2} - \frac{1}{r} \frac{dR}{dr} + \lambda^2 \left( \frac{r^2}{r_0^2} - 1 \right) R = 0. \tag{3.2}$$

Together with the Dirichlet boundary conditions

$$R(r_1)=0, \qquad R(r_2)=0,$$

equation (3.2) defines the radial spectral problem.

Multiplying equation (3.2) by $1/r$, we obtain the equivalent divergence-form representation

$$-\frac{d}{dr}\left(\frac{1}{r}\frac{dR}{dr}\right)=\lambda^2\frac{1}{r}\left(\frac{r^2}{r_0^2}-1\right)R.$$

Equivalently,

$$-\frac{d}{dr}\left(p(r)\frac{dR}{dr}\right)=\lambda^2\rho(r)R, \tag{3.3}$$

where

$$p(r)=\frac{1}{r}, \qquad \rho(r)=\frac{1}{r}\left(\frac{r^2}{r_0^2}-1\right) \tag{3.4}$$

denote the Sturm–Liouville coefficients.

Since $r_1>r_0$, the weight function satisfies $\rho(r)>0$, $r\in[r_1,r_2]$. Moreover,

$$p,\rho\in C^1([r_1,r_2]), \qquad p(r)>0, \quad \rho(r)>0,$$

on the closed finite interval. Consequently, equation (3.3) defines a regular Sturm–Liouville spectral problem with separated boundary conditions.

We introduce the weighted Hilbert space $\mathcal{H}=L^2_\rho(r_1,r_2)$, equipped with the scalar product

$$\langle f,g\rangle_\rho=\int_{r_1}^{r_2} f(r)\overline{g(r)}\rho(r)dr.$$

The associated radial operator is defined by

$$\boldsymbol{\mathcal{L}}=-\frac{1}{\rho(r)}\frac{d}{dr}\left(p(r)\frac{d}{dr}\right)$$

on the domain

$$D(\boldsymbol{\mathcal{L}})=\left(R\in H^1_0(r_1,r_2):\frac{d}{dr}\left(p(r)\frac{dR}{dr}\right)\in L^2_\rho(r_1,r_2)\right).$$

**Proposition 3.1 (Regular self-adjoint spectral structure).** *The operator $\boldsymbol{\mathcal{L}}$ is self-adjoint in $\mathcal{H}$. Its spectrum is purely discrete and consists of a sequence of positive eigenvalues*

$$0 < \lambda_1^2 \le \lambda_2^2 \le \ldots, \qquad \lambda_n^2 \to +\infty. \tag{3.5}$$

*The corresponding eigenfunctions form a complete orthonormal basis in* $\mathcal{H}$.

*Proof.* The coefficients $p(r)$ and $\rho(r)$ are continuously differentiable and strictly positive on the closed interval $[r_1, r_2]$, while the boundary conditions are regular and separated. Hence, the operator $\mathcal{L}$ defines a regular Sturm–Liouville realization in the weighted Hilbert space $\mathcal{H}$.

The associated boundary form is given by

$$p(r)\left(u(r)\overline{v'(r)} - \overline{u'(r)}v(r)\right)\Big|_{r_1}^{r_2}.$$

It vanishes identically under the Dirichlet boundary conditions, which implies the symmetry of $\mathcal{L}$.

The self-adjointness, discreteness of the spectrum, orthogonality, and completeness of the eigenfunctions are standard consequences of regular Sturm–Liouville theory (see, for example, [1,2,12]). The positivity of the eigenvalues follows from the positivity of the quadratic form

$$\int_{r_1}^{r_2} p(r)\left|R'(r)\right|^2 dr$$

under homogeneous Dirichlet boundary conditions.

Consequently, the eigenfunctions of $\mathcal{L}$ generate an exact orthogonal spectral decomposition in the weighted Hilbert space $\mathcal{H}$ [4,5].

This Sturm–Liouville framework provides the operator foundation for the exact spectral representation and the semiclassical asymptotic analysis developed in the subsequent sections.

## 4. Exact Spectral Representation

Let $\{R_m\}_{m=1}^{\infty}$ be the complete orthonormal basis of eigenfunctions of the regular Sturm–Liouville problem (3.3) in the weighted Hilbert space $\mathcal{H} = L_\rho^2(r_1, r_2)$ with scalar product

$$\langle f, g \rangle_\rho = \int_{r_1}^{r_2} \rho(r) f(r)\overline{g(r)}\, dr. \tag{4.1}$$

The eigenfunctions satisfy

$$\mathcal{L}R_m = \lambda_m^2 R_m, \qquad R_m(r_1) = R_m(r_2) = 0. \tag{4.2}$$

Let the initial data be expanded as

$$\varphi(r)=\sum_{m=1}^{\infty}a_m R_m(r), \qquad \psi(r)=\sum_{m=1}^{\infty}b_m R_m(r), \tag{4.3}$$

where

$$a_m=\langle \varphi, R_m\rangle_\rho, \qquad b_m=\langle \psi, R_m\rangle_\rho. \tag{4.4}$$

We seek the solution in the exact spectral form

$$u(r,t)=\sum_{m=1}^{\infty}c_m(t)R_m(r). \tag{4.5}$$

Substitution of the spectral expansion (4.5) into the evolution equation (2.1) and projection onto the eigenfunctions $R_m$ yield the decoupled modal equations

$$\frac{d^2c_m(t)}{dt^2}+c^2\lambda_m^2c_m(t)=0. \tag{4.6}$$

Introducing the modal frequencies $\omega_m=c\lambda_m$, we obtain

$$c_m(t)=a_m\cos(\omega_m t)+\frac{b_m}{\omega_m}\sin(\omega_m t). \tag{4.7}$$

Consequently, the exact spectral representation of the solution is

$$u(r,t)=\sum_{m=1}^{\infty}\left(a_m\cos(\omega_m t)+\frac{b_m}{\omega_m}\sin(\omega_m t)\right)R_m(r). \tag{4.8}$$

This representation shows that the evolution is completely determined by the exact eigenfrequencies $\omega_n$ and the corresponding Sturm–Liouville eigenfunctions. The exact spectral data therefore determine the modal organization of the radial dynamics.

The corresponding conserved energy is

$$E(t)=\frac{1}{2}\int_{r_1}^{r_2}\left(\frac{\rho(r)}{c^2}|u_t(r,t)|^2+p(r)|u_r(r,t)|^2\right)dr. \tag{4.9}$$

Using orthogonality of the eigenfunctions, the energy admits the spectral representation

$$E(t)=\frac{1}{2}\sum_{m=1}^{\infty}\left(\frac{1}{c^2}\left|\frac{dc_m(t)}{dt}\right|^2+\lambda_m^2|c_m(t)|^2\right). \tag{4.10}$$

In particular,

$$E(t) = E(0), \tag{4.11}$$

which follows either from (4.6) or directly by differentiating (4.9) and using the Dirichlet boundary conditions.

*4.1. Modal expansion using quasimodes*. In the exact spectral representation (4.8) the solution is expanded in the orthonormal basis of exact eigenfunctions $\{R_m\}_{m=1}^{\infty}$. However, these eigenfunctions are not known in closed form. Instead, we have constructed the explicit Liouville–Green quasimodes

$$\hat{R}_m^{(LG)}(r) = \sqrt{\frac{2}{L}}\, R_m^{(LG)}(r) = \sqrt{\frac{2}{L}}\,\frac{1}{\left(p(r)\rho(r)\right)^{1/4}}\sin\left(\frac{\pi m}{L}\xi(r)\right), \tag{4.12}$$

which, as proved in Appendix A, form an orthonormal basis of $\mathcal{H} = L_\rho^2(r_1, r_2)$. Therefore, any function $\varphi \in \mathcal{H}$ admits a Fourier expansion

$$\varphi(r) = \sum_{m=1}^{\infty} \hat{a}_m \hat{R}_m^{(LG)}(r), \qquad \hat{a}_m = \left\langle \varphi, \hat{R}_m^{(LG)} \right\rangle_\rho, \tag{4.13}$$

with convergence in the norm of $\mathcal{H}$. Applying this to the initial data $\varphi(r)$ and $\psi(r)$, we set

$$\hat{a}_m = \left\langle \varphi, \hat{R}_m^{(LG)} \right\rangle_\rho, \qquad \hat{b}_m = \left\langle \psi, \hat{R}_m^{(LG)} \right\rangle_\rho, \tag{4.14}$$

and define $\tilde{\lambda}_m = \dfrac{\pi m}{L}$, $\tilde{\omega}_m = c\tilde{\lambda}_m$. Then the series

$$u_{\text{app}}(r,t) = \sum_{m=1}^{\infty} \left( \hat{a}_m \cos(\tilde{\omega}_m t) + \frac{\hat{b}_m}{\tilde{\omega}_m} \sin(\tilde{\omega}_m t) \right) \hat{R}_m^{(LG)}(r) \tag{4.15}$$

converges in $\mathcal{H}$ for each fixed $t$ (by Parseval's identity) and represents a finite-energy solution in the weak sense.

We will consider uniform convergence for smooth initial data. Assume that $\varphi(r)$ and $\psi(r)$ satisfy the compatibility conditions

$$\varphi(r_i) = \psi(r_i) = 0, \qquad \varphi''(r_i) - \frac{1}{r_i}\varphi'(r_i) = 0 \qquad (i = 1, 2), \tag{4.16}$$

and smoothness conditions

$$\varphi \in C^3[r_1, r_2], \quad \varphi^{(4)} \in L^2_\rho(r_1, r_2), \qquad \psi \in C^1[r_1, r_2], \quad \psi^{(2)} \in L^2_\rho(r_1, r_2). \tag{4.17}$$

Then the series (4.15) converges uniformly on $[r_1, r_2] \times [0, T]$ for any $T > 0$; it can be differentiated termwise twice, and the limit function satisfies the wave equation (2.1), the boundary conditions (2.2) and the initial conditions (2.3) in the classical sense. The proof is given in Appendix B.

**Proposition 4.1 (Spectral energy representation).** *Assume that* $\varphi \in D(\mathcal{L}^{1/2})$ *and* $\psi \in \mathcal{H}$. *Then the series (4.8) defines a unique finite-energy solution for admissible initial data. Moreover, the energy* $E(t)$ *is conserved and is given by the spectral identity (4.10).*

This exact spectral representation provides the starting point for the asymptotic analysis developed in the subsequent sections. The Liouville–Green approximation constructed in Sects. 5 and 6 yields explicit asymptotic representations of the spectral modes and thereby enables an explicit approximation of the exact solution.

*Remark 4.1.* The Sturm–Liouville coefficient pair $p(r), \rho(r)$ defined in (3.4) plays a distinguished role throughout the analysis. Besides determining the self-adjoint spectral operator, it simultaneously generates the variational form, the weighted Hilbert space structure, and the conserved energy representation of the evolution problem. The same coefficient pair also gives rise to the logarithmic Liouville coordinate governing the high-mode asymptotic behaviour of the spectrum.

## 5. Liouville–Green High-Frequency Asymptotics

We consider the regular Sturm–Liouville problem (3.3) on the finite interval $[r_1, r_2]$, where the coefficients $p(r)$ and $\rho(r)$ are defined by (3.4). By Assumption 2.1, the weight function satisfies $\rho(r) > 0$ on $[r_1, r_2]$. Therefore, the problem admits the standard Liouville transformation and the classical Liouville–Green asymptotic analysis applies.

*5.1. Liouville transformation.* Introduce the Liouville variable

$$\xi(r) = \int_{r_1}^{r} \sqrt{\frac{\rho(s)}{p(s)}}\, ds, \qquad 0 \le \xi \le L, \tag{5.1}$$

where

$$L = \xi(r_2) = \int_{r_1}^{r_2} \sqrt{\frac{\rho(s)}{p(s)}}\, ds\,. \tag{5.2}$$

Since the coefficients $p(r)$ and $\rho(r)$ are defined by (3.4), we have

$$\frac{\rho(r)}{p(r)}=\frac{r^2}{r_0^2}-1.$$

Therefore, by the geometric assumption $r_1, r_2 > r_0$, the Liouville transformation associated with the equation is regular on the whole interval [6,7,13]. In the present case,

$$L=\int_{r_1}^{r_2}\sqrt{\frac{r^2}{r_0^2}-1}\,dr,$$

and hence

$$L=\frac{1}{2}\left(r\sqrt{\frac{r^2}{r_0^2}-1}-r_0\ln\frac{r+\sqrt{r^2-r_0^2}}{r_0}\right)\Bigg|_{r_1}^{r_2}. \tag{5.3}$$

Let us define

$$U(\xi)=\left(p(r)\rho(r)\right)^{1/4}R(r).$$

Then the Sturm–Liouville equation is reduced to the normal Liouville form

$$-\frac{d^2U(\xi)}{d\xi^2}+V(\xi)U(\xi)=\lambda^2U(\xi), \qquad \xi\in[0,L], \tag{5.4}$$

where $V\in C^{\infty}\left([0,L]\right)$. The Dirichlet boundary conditions for $R$ are transformed into

$$U(0)=U(L)=0.$$

*5.2. Spectral asymptotics.* Since the transformed equation is posed on a finite interval, contains no turning points, and possesses a smooth potential $V$, the standard Liouville–Green theory yields

$$\lambda_m=\frac{m\pi}{L}+O(m^{-1}), \qquad m\to\infty. \tag{5.5}$$

Moreover, the corresponding eigenfunctions admit the leading asymptotic representation

$$R_m(r)\sim\frac{1}{\left(p(r)\rho(r)\right)^{1/4}}\sin\left(\lambda_m\int_{r_1}^{r}\sqrt{\frac{\rho(s)}{p(s)}}\,ds\right), \qquad m\to\infty, \tag{5.6}$$

uniformly for $r \in [r_1, r_2]$, up to normalization.

The asymptotic formulae (5.5)–(5.6) are standard consequences of the Liouville–Green method for regular Sturm–Liouville problems without turning points. Rigorous error estimates and higher-order asymptotic expansions follow from the classical theory of Olver [6] and related modern developments in spectral asymptotics [5].

*5.3. Explicit Liouville–Green approximations of the radial eigenpairs*. We now specialize the general Liouville–Green asymptotics obtained in Sections 5.1 and 5.2 to the coefficients $p(r)$ and $\rho(r)$ defined in (3.4), and derive explicit closed-form approximations for the high-mode eigenpairs.

Recall the Liouville variable introduced in (5.1):

$$\xi(r) = \int_{r_1}^{r} \sqrt{\frac{s^2}{r_0^2} - 1}\, ds.$$

A direct integration yields the alternative representation

$$\xi(r) = w(r) - w(r_1),$$

where

$$w(r) = \frac{1}{2}\left( r\sqrt{\frac{r^2}{r_0^2} - 1} - r_0 \ln \frac{r + \sqrt{r^2 - r_0^2}}{r_0} \right). \tag{5.7}$$

At leading asymptotic order, the Dirichlet boundary conditions $R(r_1) = R(r_2) = 0$ lead to the quantization condition

$$\lambda_m L \sim \pi m, \qquad m \to \infty.$$

Hence the corresponding explicit Liouville–Green approximations of the eigenvalues are given by

$$\tilde{\lambda}_m = \frac{\pi m}{L}. \tag{5.8}$$

Substituting (5.8) into the leading-order asymptotic expression (5.6) and choosing the phase so that the sine vanishes at $r = r_1$, we obtain the explicit Liouville–Green quasimodes

$$R_m^{(LG)}(r) = \frac{1}{\left(p(r)\rho(r)\right)^{1/4}} \sin\left( \frac{\pi m}{L} \xi(r) \right). \tag{5.9}$$

By construction,

$$R_m^{(LG)}(r_1)=0, \qquad R_m^{(LG)}(r_2)=0.$$

**Proposition 5.1.** *The explicit Liouville–Green quasimodes* $R_m^{(LG)}$ *defined by (5.9) are orthogonal with respect to the weight* $\rho(r)$:

$$\int_{r_1}^{r_2} \rho(r) R_m^{(LG)} R_k^{(LG)} dr = 0, \qquad m \neq k. \tag{5.10}$$

*A direct computation using the Liouville transformation yields the normalization constant* $\left\| R_m^{(LG)} \right\|_\rho = \sqrt{L/2}$. *Hence the normalized quasimodes are* $\hat{R}_m^{(LG)} = \sqrt{2/L}\, R_m^{(LG)}$, *and they form an orthonormal system in the weighted Hilbert space* $\mathcal{H} = L_\rho^2(r_1, r_2)$.

*Proof.* Under the Liouville transformation,

$$d\xi = \sqrt{\frac{\rho(r)}{p(r)}} dr,$$

while the quasimodes (5.9) reduce to sine functions in the variable $\xi$. Therefore, the orthogonality relation (5.10) follows directly from the standard orthogonality of sine functions on $[0, L]$. This completes the proof.

The functions $R_m^{(LG)}$ should be regarded as explicit Liouville–Green quasimodes associated with the original Sturm–Liouville operator. They satisfy the boundary conditions exactly, preserve the correct high-frequency oscillatory structure, and reproduce the orthogonal structure of the exact Sturm–Liouville eigenfunctions with respect to the original weight $\rho(r)$. Because of their explicit structure, these quasimodes provide a natural approximation system for subsequent spectral representations and asymptotic analysis. Their quantitative relation to the exact spectral modes and the corresponding asymptotic spectral representation will be analyzed in Sects. 6 and 7.

## 6. Asymptotic Orthogonality and Spectral Approximation of Liouville–Green Quasimodes

Throughout this section we assume $r_1, r_2 > r_0$, with $r_1 < r_2$.

Let $\{R_m\}_{m=1}^{\infty}$ be the orthonormal eigenfunctions of the Sturm–Liouville operator $\mathcal{L}$ introduced in Section 4, corresponding to the eigenvalues $\lambda_m^2$. Recall that the eigenfunctions form a complete orthonormal basis in the weighted Hilbert space $\mathcal{H} = L^2_\rho(r_1, r_2)$ with scalar product

$$\langle f, g \rangle_\rho = \int_{r_1}^{r_2} \rho(r) f(r) \overline{g(r)} dr.$$

The purpose of this section is to compare the exact spectral modes $R_m$ with the normalized Liouville–Green quasimodes $\hat{R}_m^{(LG)}$ constructed in Section 5.3 and to quantify their asymptotic proximity in the high-frequency regime.

*6.1. Liouville–Green approximation of the exact eigenfunctions.* Recall the Liouville variable introduced in (5.1):

$$\xi(r) = \int_{r_1}^{r} \sqrt{\frac{\rho(s)}{p(s)}} ds, \qquad 0 \le \xi \le L. \tag{6.1}$$

By the Liouville–Green asymptotic expansion (5.6) and the normalization (4.12), the exact eigenfunctions admit the representation

$$R_m(r) = \hat{R}_m^{(LG)}(r) + \varepsilon_m(r), \tag{6.2}$$

where the remainder satisfies

$$\varepsilon_m(r) = O(m^{-1}), \qquad m \to \infty,$$

uniformly [6,7] for $r \in [r_1, r_2]$. The eigenvalues satisfy the classical asymptotic relation

$$\lambda_m = \tilde{\lambda}_m + O(m^{-1}), \qquad \tilde{\lambda}_m = \frac{\pi m}{L}, \qquad m \to \infty. \tag{6.3}$$

Consequently, the exact spectral modes $R_m$ are asymptotically approximated by the normalized Liouville–Green quasimodes

$$\hat{R}_m^{(LG)}(r) = \sqrt{\frac{2}{L}} \frac{1}{\left(p(r)\rho(r)\right)^{1/4}} \sin\left(\tilde{\lambda}_m \xi(r)\right).$$

In particular, combining (6.2) with (6.3) we obtain

$$R_m(r) - \hat{R}_m^{(LG)}(r) = O(m^{-1}), \qquad m \to \infty, \tag{6.4}$$

uniformly on $[r_1, r_2]$. Thus, the explicit Liouville–Green quasimodes reproduce the leading high-frequency structure of the exact Sturm–Liouville eigenfunctions and provide an explicit asymptotic model for the spectral modes of the operator $\mathcal{L}$.

*6.2. Off-diagonal scalar products.* To analyze the asymptotic interaction, consider the mixed off-diagonal scalar products $\left\langle R_m, \hat{R}_k^{(LG)} \right\rangle_\rho$ with $m \neq k$.

We first compute the scalar products of the quasimodes themselves. Passing to the Liouville variable $\xi$,

$$\left\langle \hat{R}_m^{(LG)}, \hat{R}_k^{(LG)} \right\rangle_\rho = \int_0^L \sin(\tilde{\lambda}_m \xi) \sin(\tilde{\lambda}_k \xi) d\xi = \delta_{mk}, \tag{6.5}$$

since $\tilde{\lambda}_m = \pi m / L$ and the functions $\sqrt{2/L} \sin(\pi m \xi / L)$ are orthonormal on $[0, L]$. Hence $\left\{ \hat{R}_m^{(LG)} \right\}$ is an orthonormal system in $\mathcal{H}$ (which is already known from Appendix A).

Using the representation (6.2) together with the orthogonality (6.5), we obtain for $m \neq k$ that

$$\left\langle R_m, \hat{R}_k^{(LG)} \right\rangle_\rho = \left\langle \hat{R}_m^{(LG)}, \hat{R}_k^{(LG)} \right\rangle_\rho + \left\langle \varepsilon_m, \hat{R}_k^{(LG)} \right\rangle_\rho = 0 + O(m^{-1}). \tag{6.6}$$

Thus, the off-diagonal interaction between exact eigenfunctions and quasimodes becomes asymptotically negligible.

*6.3. Normalization and spectral approximation in the Hilbert space.* Estimate (6.4) directly implies

$$\left\| R_m - \hat{R}_m^{(LG)} \right\|_\rho = O(m^{-1}), \qquad m \to \infty. \tag{6.7}$$

Since the exact eigenfunctions $\{R_k\}_{k=1}^{\infty}$ form a complete orthonormal basis in $\mathcal{H}$, every quasimode admits the spectral expansion

$$\hat{R}_m^{(LG)} = \sum_{k=1}^{\infty} c_{mk} R_k, \qquad c_{mk} = \left\langle \hat{R}_m^{(LG)}, R_k \right\rangle_\rho.$$

Using (6.7) and Parseval's identity we find

$$c_{mm} = 1 + O(m^{-1}), \qquad c_{mk} = O(m^{-1}), \qquad k \neq m. \tag{6.8}$$

Hence the normalized Liouville–Green quasimodes become asymptotically aligned with the corresponding exact spectral modes of $\mathcal{L}$.

As a consequence of (6.8), for any function $\phi$ satisfying the smoothness conditions (4.17) we obtain the asymptotic relation for its Fourier coefficients

$$a_m - \hat{a}_m = O(m^{-1}), \qquad b_m - \hat{b}_m = O(m^{-1}), \qquad m \to \infty.$$

Indeed, expanding $\hat{a}_m = \sum_k c_{mk} a_k$ and using $a_k = O(k^{-2})$ together with the estimates for $c_{mk}$ yields the result. The same argument applies to $b_m$.

The estimates obtained in this section establish a quantitative connection between the explicit quasimodes and the exact spectral structure. They provide the asymptotic framework for the solution expansion developed in Section 4 and for the error analysis in Section 7.

## 7. Spectral Dynamics and Semiclassical Asymptotics

In this section we combine the exact spectral representation of Section 4 with the asymptotic analysis of Sects. 5 and 6. We show how the explicit Liouville–Green quasimodes approximate the exact solution and quantify the error.

Throughout this section we assume Assumption 2.1 and the Dirichlet boundary conditions on $[r_1, r_2]$, with $r_1, r_2 > r_0$. The weight function

$$\rho(r) = \frac{r}{r_0^2} - \frac{1}{r}$$

is strictly positive on $[r_1, r_2]$, and the Sturm–Liouville operator $\mathcal{L}$ is self-adjoint in $\mathcal{H} = L_\rho^2(r_1, r_2)$.

*7.1. Spectral representation of the solution.* Let $\{R_m\}_{m=1}^{\infty}$ be the complete orthonormal basis of eigenfunctions $\mathcal{L}$ with eigenvalues $\lambda_m^2$. By the spectral theorem, every admissible solution of (2.1)–(2.3) admits the expansion (4.5)–(4.8):

$$u_{\text{exact}}(r,t) = \sum_{m=1}^{\infty} \left( a_m \cos(\omega_m t) + \frac{b_m}{\omega_m} \sin(\omega_m t) \right) R_m(r), \qquad \omega_m = c\lambda_m,$$

where $a_m = \langle \varphi, R_m \rangle_\rho$, $b_m = \langle \psi, R_m \rangle_\rho$.

The exact eigenfunctions and eigenvalues are not known in closed form. However, using the Liouville–Green method we have constructed explicit approximations $\tilde{\lambda}_m$ and $\hat{R}_m^{(LG)}$ given by (5.8) and (4.12). They satisfy: $\tilde{\lambda}_m = \pi m/L$ with $L$ from (5.3); $\{\hat{R}_m^{(LG)}\}$ is an orthonormal basis of $\mathcal{H}$ (Appendix A). For each $m$, the Fourier coefficients of the initial data with respect to this basis are $\hat{a}_m = \langle \varphi, \hat{R}_m^{(LG)} \rangle_\rho$, $\hat{b}_m = \langle \psi, \hat{R}_m^{(LG)} \rangle_\rho$.

The approximate solution is defined by (4.15):

$$u_{\text{app}}(r,t) = \sum_{m=1}^{\infty} \left( \hat{a}_m \cos(\tilde{\omega}_m t) + \frac{\hat{b}_m}{\tilde{\omega}_m} \sin(\tilde{\omega}_m t) \right) \hat{R}_m^{(LG)}(r), \qquad \tilde{\omega}_m = c\tilde{\lambda}_m.$$

*7.2. Asymptotic closeness of the eigenfunctions.* From the Liouville–Green asymptotics (5.5)–(5.6) and the explicit expressions (5.8)–(5.9), the following estimates are proved in Section 6:

$$\lambda_m = \tilde{\lambda}_m + O(m^{-1}),$$

$$R_m(r) = R_m^{(LG)}(r) + O(m^{-1}) \quad \text{uniformly on } [r_1, r_2],$$

$$\left\| R_m - \hat{R}_m^{(LG)} \right\|_\rho = O(m^{-1}).$$

These estimates show that the quasimodes capture the leading oscillatory behaviour of the exact eigenfunctions.

*7.3. Error estimate for the solution.*

**Lemma 7.1 (Error estimate).** *Assume that the initial data* $\varphi, \psi$ *satisfy the smoothness conditions (4.17). Then for any* $t \geq 0$

$$\left\| u_{\text{exact}}(r,t) - u_{\text{app}}(r,t) \right\|_\rho \leq C(t) \sum_{m=1}^{\infty} \left( |\hat{a}_m| + |\hat{b}_m| \right) \left\| R_m - \hat{R}_m^{(LG)} \right\|_\rho$$

*where* $C(t) = \max(1, 1/c)(1+t)$. *Using the asymptotic estimates (6.7) and the fact that* $|\hat{a}_m|, |\hat{b}_m| = O(m^{-2})$ *(which follows from (4.17) by integration by parts), we obtain*

$$\left\| u_{\text{exact}}(r,t) - u_{\text{app}}(r,t) \right\|_\rho = O(t) \sum_{m=1}^{\infty} \frac{1}{m^3} = O(t).$$

*In particular, for any fixed* $t$ *the error is finite, and the modal approximation error decreases as* $m \to \infty$. *The convergence is uniform on bounded time intervals.*

*Proof.* The difference $w = u_{\text{exact}} - u_{\text{app}}$ satisfies the wave equation with initial data given by the differences of the series. Using the triangle inequality, orthonormality of the two bases, and the mean value theorem for trigonometric functions together with $|\omega_m - \tilde{\omega}_m| = O(m^{-1})$ from (6.3), we obtain the stated bound. The decay of $\hat{a}_m, \hat{b}_m$ follows from integration by parts under conditions (4.17). The details are standard and are omitted here.

*7.4. Asymptotic orthogonality and off-diagonal interactions.* For the mixed scalar products $\left\langle R_m, R_k^{(LG)} \right\rangle_\rho$ with $m \neq n$, estimate (6.6) gives

$$\left\langle R_m, R_k^{(LG)} \right\rangle_\rho = O(m^{-1}), \qquad m \to \infty.$$

Thus, the interaction between exact eigenfunctions and quasimodes becomes negligible for large mode numbers. A sharper estimate can be obtained using Lemma C.1 on the nonstationary phase (see Appendix C), which shows that

$$\left\langle R_m^{(LG)}, R_k^{(LG)} \right\rangle_\rho = O(|m-n|^{-1}), \qquad |m-n| \to \infty,$$

and consequently the off-diagonal contributions in the spectral expansions vanish asymptotically. This justifies the asymptotic orthogonality of the quasimode system.

*7.5. Summary.* The analysis of this section demonstrates that the explicit Liouville–Green quasimodes provide a practical and asymptotically accurate approximation to the exact spectral solution of the radial wave equation. The estimates (6.4), (6.7) and Lemma 7.1 quantify the approximation error, while the results of Appendix C explain the asymptotic decay of off-diagonal terms. The overall framework provides a consistent connection between the exact Sturm–Liouville formulation and the Liouville–Green asymptotic description.

## 8. Numerical Verification of the Liouville–Green Spectrum

To assess the accuracy of the Liouville–Green approximation derived in Section 5, we compare the asymptotic eigenvalue formula (5.8) with numerical solutions of the exact Sturm–Liouville problem (3.2)–(3.3). For the representative parameter values

$$r_0 = 1, \qquad r_1 = 5, \qquad r_2 = 10,$$

the eigenvalues of the exact boundary-value problem were computed numerically.

The numerical computation was performed directly for the radial equation (3.2) with Dirichlet boundary conditions at $r = r_1$ and $r = r_2$. The interval $[r_1, r_2]$ was discretized and the resulting self-adjoint Sturm–Liouville eigenvalue problem was solved numerically. The computed eigenvalues were then compared with the Liouville–Green prediction

$$\tilde{\lambda}_m = \frac{\pi m}{L},$$

where $L = w(r_2) - w(r_1)$ and $w(r)$ is defined by (5.7) (equivalently, $L$ is given by (5.3)).

The relative error is defined as

$$\delta_m = \frac{\left|\tilde{\lambda}_m - \lambda_m^{\text{num}}\right|}{\lambda_m^{\text{num}}} \times 100\%.$$

The comparison is summarized in Table 1.

**Table 1.** Numerical eigenvalues $\lambda_m^{\text{num}}$ and Liouville–Green approximations $\tilde{\lambda}_m$ for $r_0 = 1,\ r_1 = 5,\ r_2 = 10$

| $m$ | $\tilde{\lambda}_m$ | $\lambda_m^{\text{num}}$ | $\delta_m (\%)$ |
|---|---|---|---|
| 1 | 0.08456 | 0.08449 | 0.08 |
| 2 | 0.16912 | 0.16907 | 0.03 |
| 3 | 0.25368 | 0.25364 | 0.02 |
| 4 | 0.33826 | 0.33820 | 0.02 |
| 5 | 0.42280 | 0.42276 | 0.01 |
| 6 | 0.50737 | 0.50732 | 0.01 |
| 7 | 0.59194 | 0.59186 | 0.01 |
| 8 | 0.67648 | 0.67639 | 0.01 |
| 9 | 0.76105 | 0.76093 | 0.02 |
| 10 | 0.84559 | 0.84545 | 0.02 |
| 11 | 0.93016 | 0.92996 | 0.02 |
| 12 | 1.0147 | 1.0144 | 0.03 |

| 13 | 1.0993 | 1.0990 | 0.03 |
|---|---|---|---|
| 14 | 1.1838 | 1.1834 | 0.03 |
| 15 | 1.2684 | 1.2679 | 0.04 |
| 16 | 1.3530 | 1.3523 | 0.05 |
| 17 | 1.4375 | 1.4368 | 0.05 |
| 18 | 1.5221 | 1.5212 | 0.06 |
| 19 | 1.6066 | 1.6056 | 0.06 |
| 20 | 1.6912 | 1.6900 | 0.07 |

The numerical eigenvalues were computed from the exact self-adjoint Sturm–Liouville formulation (3.3) using a finite-difference discretization [14]. Grid refinement tests confirmed the stability of the reported eigenvalues. The relative errors remain below $0.08\,\%$ for all computed modes and are typically of the order of $0.01\,\%$. This behaviour is consistent with the theoretical estimate $\lambda_m - \tilde{\lambda}_m = O(m^{-1})$.

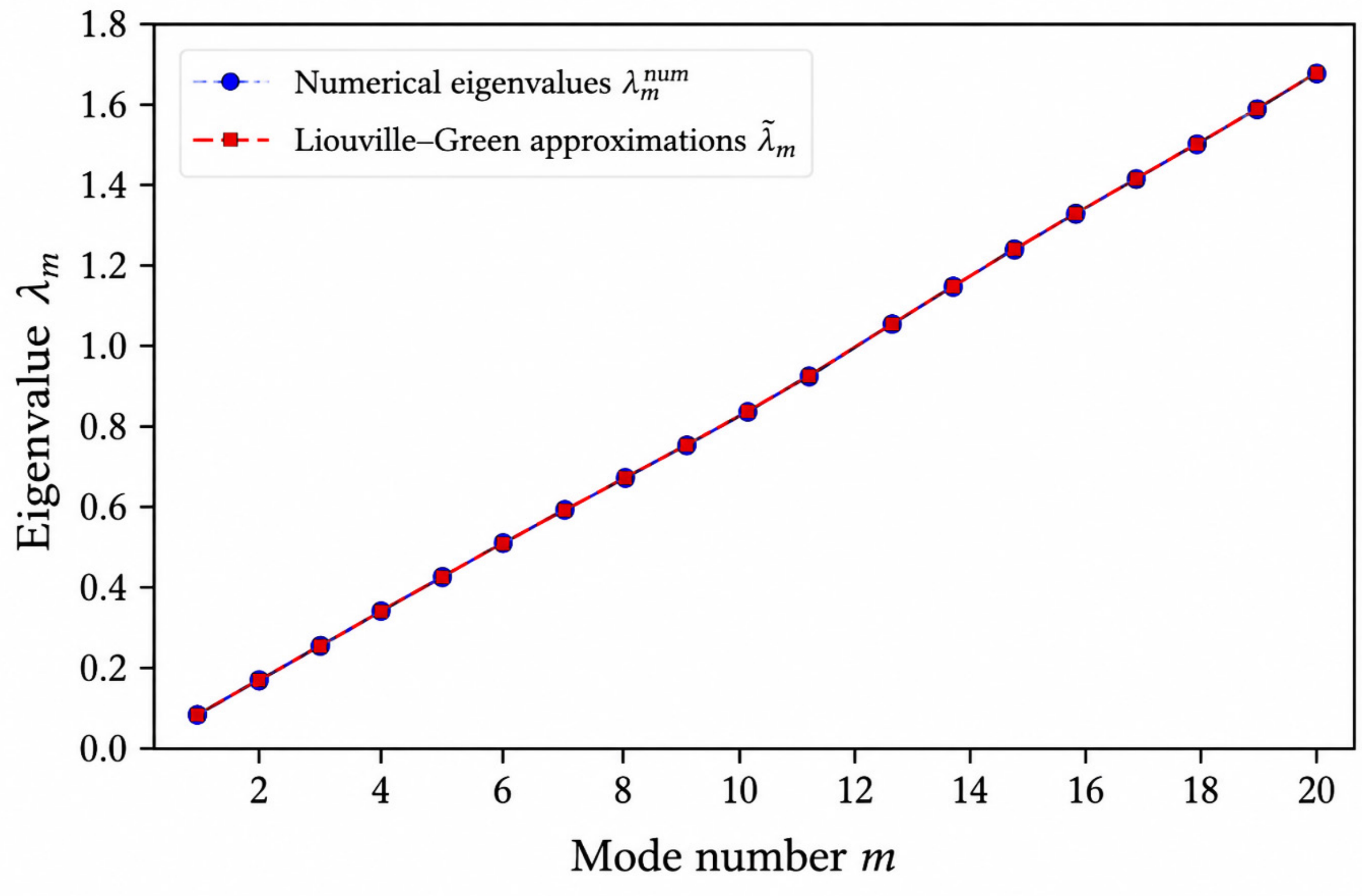

**Fig. 1.** Comparison of the exact numerical eigenvalues $\lambda_m^{\text{num}}$ and Liouville–Green approximations $\tilde{\lambda}_m$ for the first ten modes

The numerical eigenvalues and their Liouville–Green approximations are shown in Fig. 1. The two curves are visually indistinguishable at the scale of the figure, while the relative errors reported in Table 1 remain below $0.08\,\%$ for all computed modes.

These numerical results provide independent validation of the asymptotic quantization formula (5.8) derived in Section 5. Combined with the asymptotic reconstruction developed in Section 7 and the spectral properties established in Section 6, they demonstrate that the Liouville–Green approximation accurately captures both the spectral characteristics of the variable-coefficient radial Sturm–Liouville operator and the corresponding reconstruction of solutions to the original boundary-value problem.

## 9. Conclusions

We have developed a complete spectral analysis of the radial Sturm–Liouville operator with variable coefficients arising from a radially symmetric wave equation whose singular point lies outside the interval under consideration. The original initial-boundary value problem was reduced to a regular self-adjoint Sturm–Liouville problem on a finite interval, thereby establishing the exact spectral framework governing the associated radial evolution.

The resulting variable-coefficient radial operator was shown to possess a discrete real spectrum and a complete orthonormal system of eigenfunctions in the appropriate weighted Hilbert space. This yields an exact spectral representation of the evolution problem and provides the foundation for the subsequent asymptotic analysis.

An exact transcendental equation governing the spectrum was derived from the boundary conditions. To investigate the high-frequency regime, the Liouville–Green transformation was applied directly to the radial Sturm–Liouville equation, leading to explicit asymptotic formulas for the eigenvalues and eigenfunctions. The resulting quasimodes were shown to remain compatible with the weighted spectral structure of the exact operator and to asymptotically reproduce the corresponding orthogonality relations.

These asymptotic spectral data were further incorporated into the reconstruction of solutions of the original boundary-value problem. The completeness of the quasimode system and the convergence properties of the associated modal expansions were established, while oscillatory integral estimates provided quantitative control of off-diagonal spectral interactions in the high-mode regime.

Numerical computations of the exact Sturm–Liouville spectrum demonstrated excellent agreement with the Liouville–Green predictions. For all computed modes, the relative errors remained below $0.08\,\%$, thereby providing independent validation of the asymptotic quantization formula and confirming the accuracy of the proposed spectral approximation.

More broadly, the present work connects the exact spectral formulation of the radial Sturm–

Liouville problem with its asymptotic Liouville–Green description, the corresponding reconstruction of solutions, and independent numerical verification.

The results show that the Liouville–Green approximation provides an accurate and spectrally consistent description of the considered radial operator in the high-frequency regime. The framework developed here may serve as a basis for further investigations of more general variable-coefficient radial operators, perturbative spectral problems, and related models arising in mathematical physics.

**Acknowledgements.** I.V.G. acknowledges support by the National Research Foundation of Ukraine under Grant No. 2025.07/0132.

**Declarations**

**Conflict of interest.** The authors have no competing interests to declare that are relevant to the content of this article.

## Appendix A. Completeness of the Liouville–Green Quasimode System

In this appendix we prove that the system of Liouville–Green quasimodes $\left\{R_m^{(LG)}\right\}_{m=1}^{\infty}$ defined in Section 5.3 is **complete** in the weighted Hilbert space $\mathcal{H} = L_\rho^2(r_1, r_2)$ with weight $\rho(r) = \frac{1}{r}\left(\frac{r^2}{r_0^2} - 1\right)$.

Completeness means that every element of $\mathcal{H}$ can be approximated arbitrarily closely in the weighted norm by finite linear combinations of the quasimodes. Equivalently, the linear span of the quasimodes is dense in $\mathcal{H}$.

To establish this property, it is sufficient to prove uniform approximation of continuous functions. Since the weight function $\rho(r)$ is continuous and strictly positive on the closed interval $[r_1, r_2]$, uniform approximation also implies approximation in the weighted $L_\rho^2$-norm.

*A.1. Sufficient condition for completeness.* A standard completeness criterion (see, e.g., [12]) states that if, for every continuous function $f \in C[r_1, r_2]$ and every $\varepsilon > 0$, there exists a finite linear combination of the system satisfying

$$\left\| f - S_N \right\|_{C[r_1, r_2]} < \varepsilon,$$

then the system is complete in $L_\rho^2(r_1, r_2)$.

It therefore suffices to establish uniform approximation by the Liouville–Green quasimodes.

*A.2. Transformation to a trigonometric system.* Recall from Section 5.1 the Liouville variable

$$\xi(r) = \int_{r_1}^{r} \sqrt{\frac{\rho(s)}{p(s)}}\, ds = \frac{1}{r_0} \int_{r_1}^{r} \sqrt{s^2 - r_0^2}\, ds,$$

and set

$$L = \xi(r_2), \qquad x(r) = \xi(r).$$

The function $x(r)$ is strictly increasing, $C^1$, with $x(r_1) = 0$ and $x(r_2) = L$. Its inverse $r = r(s)$ is also strictly increasing and continuously differentiable on $[0, L]$.

The quasimodes (5.9) can be written in the form

$$R_m^{(LG)}(r) = \frac{1}{\left(p(r)\rho(r)\right)^{1/4}} \sin\left(\frac{\pi m}{L}\xi(r)\right) = \sqrt[4]{\frac{r^2 r_0^2}{r^2 - r_0^2}} \sin\left(\frac{\pi m}{L} x(r)\right).$$

The constant factor $r_0^{1/2}$ is irrelevant for completeness and can be absorbed into the coefficients $C_m$. Thus, we set

$$A(r) = \sqrt[4]{\frac{r^2}{r^2 - r_0^2}},$$

so that

$$R_m^{(LG)}(r) = A(r) \sin\left(\frac{\pi m}{L} x(r)\right), \qquad x \in [0, L],$$

up to an inessential constant factor. The function $A(r)$ is smooth and bounded on $[r_1, r_2]$ because $r_1 > r_0$.

*A.3. Approximation of an arbitrary continuous function.* Let $f \in C[r_1, r_2]$, and let

$$g(\xi) = f(r(\xi)).$$

Then $g \in C[\xi_1, \xi_2]$. Extend $g$ to $[-\xi_2, \xi_2]$ by odd reflection.

The classical Fourier sine system is complete in $C[\xi_1, \xi_2]$. Therefore, for every $\delta > 0$ there exist an integer $N$ and coefficients $c_m$ such that

$$\|g - S_N\|_{C[\xi_1, \xi_2]} < \delta.$$

Returning to the radial variable gives

$$S_N(\xi(r)) = \sum_{m=1}^{N} c_m \sin \frac{\pi m\big(\xi(r)-\xi_1\big)}{\xi_2-\xi_1}.$$

Now let

$$q(r) = \sqrt{\frac{r_0}{r-r_0}}, \qquad M = \max_{r\in[r_1,r_2]} q(r).$$

Hence

$$M = \|q\|_{C[r_1,r_2]} < \infty.$$

Given an arbitrary $\varepsilon > 0$, choose

$$\delta = \frac{\varepsilon}{M}.$$

Then

$$\|g - S_N\|_{C[\xi_1,\xi_2]} < \delta.$$

Multiplication by the bounded function $q(r)$ yields

$$\left\|R - R_N^{(LG)}\right\|_{C[r_1,r_2]} = \|q(g-S_N)\|_{C[r_1,r_2]} \le M\,\|g-S_N\|_{C[\xi_1,\xi_2]} < M\delta = \varepsilon.$$

Since $\varepsilon > 0$ was arbitrary, the Liouville–Green quasimodes uniformly approximate every continuous function on $[r_1,r_2]$.

*A.4. Conclusion.* It follows that finite linear combinations of the Liouville–Green quasimodes are dense in $C[r_1,r_2]$. Since the weight function $\rho(r)$ is continuous and strictly positive, this density immediately extends to the weighted Hilbert space $\mathcal{H} = L_\rho^2(r_1,r_2)$.

Consequently, the normalized Liouville–Green quasimodes form a **complete** system in $\mathcal{H}$.

Together with the asymptotic orthogonality properties established in Section 6, this completeness provides the functional basis for the asymptotic spectral reconstruction developed in Section 7.

## Appendix B. Proof of Uniform Convergence

*B.1. Preliminaries.* Consider the series (4.15)

$$u_{\text{app}}(r,t) = \sum_{m=1}^{\infty}\left(\hat{a}_m\cos(\tilde{\omega}_m t) + \frac{\hat{b}_m}{\omega_m}\sin(\tilde{\omega}_m t)\right)\hat{R}_m^{(LG)}(r), \qquad \tilde{\omega}_m = c\tilde{\lambda}_m, \quad \tilde{\lambda}_m = \frac{\pi m}{L},$$

with coefficients $\hat{a}_m, \hat{b}_m$ defined by (4.14) and $\hat{R}_m^{(LG)}$ given by (4.12). The functions $\hat{R}_m^{(LG)}$ form an orthonormal basis of $\mathcal{H} = L^2_\rho(r_1, r_2)$ (Appendix A) and satisfy the uniform bound

$$\left|\hat{R}_m^{(LG)}(r)\right| \le M, \qquad \forall m \in \mathbb{N}, \quad r \in [r_1, r_2],$$

which follows from the explicit expression (4.12) and the positivity of $p$ and $\rho$. Consequently, the series for $u_{\text{app}}$, its time derivatives $u_t, u_{tt}$, and its spatial derivatives $u_r, u_{rr}$ (obtained by formal termwise differentiation) are majorized by

$$\sum_{m=1}^{\infty} \left( |\hat{a}_m| + |\hat{b}_m| \right) \quad \text{and} \quad \sum_{m=1}^{\infty} \tilde{\lambda}_m^2 \left( |\hat{a}_m| + |\hat{b}_m| \right),$$

up to constant factors. Thus, it suffices to prove that these two series converge.

*B.2. Decay estimates for the coefficients.* Assume that the initial data satisfy the smoothness and compatibility conditions (4.16)–(4.17).

**Estimate for $\hat{a}_m$.** Using the explicit form (4.12) of the quasimodes, we apply the operator $\mathcal{L} = -\dfrac{1}{\rho(r)} \dfrac{d}{dr}\left( p(r) \dfrac{d}{dr} \right)$ and integrate by parts four times. Because $\varphi$ vanishes at the endpoints together with its first derivative (by (4.16)), the boundary terms vanish. We obtain

$$\hat{a}_m = \frac{1}{\tilde{\lambda}_m^4} \int_{r_1}^{r_2} \Phi(r) \rho(r) \hat{R}_m^{(LG)}(r)\, dr,$$

where $\Phi(r)$ is a linear combination of derivatives of $\varphi$ up to order four (with smooth bounded coefficients). Define the Fourier coefficients of $\Phi$ with respect to the orthonormal basis $\left\{\hat{R}_m^{(LG)}\right\}$:

$$\alpha_m = \left\langle \Phi, \hat{R}_m^{(LG)} \right\rangle_\rho = \int_{r_1}^{r_2} \Phi(r) \rho(r) \hat{R}_m^{(LG)}(r)\, dr.$$

Then $\hat{\alpha}_m = \dfrac{\alpha_m}{\tilde{\lambda}_m^4}$. By Parseval's identity,

$$\sum_{m=1}^{\infty} |\alpha_m|^2 = \|\Phi\|_\rho^2 < \infty,$$

since $\varphi^{(4)} \in L^2_\rho(r_1, r_2)$ and the lower-order derivatives are continuous. Hence $\{\alpha_m\} \in \ell^2$.

**Estimate for $\hat{b}_m$.** Similarly, two integrations by parts (using the vanishing of $\psi$ at the endpoints) yield

$$\hat{b}_m = \frac{1}{c\tilde{\lambda}_m^3} \int_{r_1}^{r_2} \Psi(r)\rho(r)\hat{R}_m^{(LG)}(r)\,dr,$$

with $\Psi(r)$ a combination of $\psi$ and $\psi'$. Define

$$\beta_m = \left\langle \Psi, \hat{R}_m^{(LG)} \right\rangle_\rho = \int_{r_1}^{r_2} \Psi(r)\rho(r)\hat{R}_m^{(LG)}(r)\,dr.$$

Then $\hat{b}_m = \dfrac{\beta_m}{c\tilde{\lambda}_m^3}$. The condition $\psi^{(2)} \in L^2_\rho(r_1, r_2)$ guarantees $\{\beta_m\} \in \ell^2$.

*B.3. Convergence of the majorizing series.* Because $\tilde{\lambda}_m \sim \pi m/L$, there exists a constant $C > 0$ such that $\tilde{\lambda}_m \geq Cm$ for all $m$. Therefore

$$\sum_{m=1}^{\infty} \left( |\hat{a}_m| + |\hat{b}_m| \right) \leq C_1 \sum_{m=1}^{\infty} \frac{1}{\tilde{\lambda}_m^3} \left( |\alpha_m| + |\beta_m| \right),$$

$$\sum_{m=1}^{\infty} \tilde{\lambda}_m^2 \left( |\hat{a}_m| + |\hat{b}_m| \right) \leq C_2 \sum_{m=1}^{\infty} \frac{1}{\tilde{\lambda}_m} \left( |\alpha_m| + |\beta_m| \right).$$

Both series converge by the Cauchy–Schwarz inequality, since $\left\{ \dfrac{1}{\tilde{\lambda}_m} \right\} \in \ell^2$ and $\{ |\alpha_m| + |\beta_m| \} \in \ell^2$.

Indeed,

$$\sum_{m=1}^{\infty} \frac{1}{\tilde{\lambda}_m} \left( |\alpha_m| + |\beta_m| \right) \leq \left( \sum_{m=1}^{\infty} \frac{1}{\tilde{\lambda}_m^2} \right)^{1/2} \left( \sum_{m=1}^{\infty} \left( |\alpha_m| + |\beta_m| \right)^2 \right)^{1/2} < \infty,$$

and for the first series (with $1/\tilde{\lambda}_m^3$) we use $\sum 1/\tilde{\lambda}_m^6 < \infty$:

$$\sum_{m=1}^{\infty} \frac{1}{\tilde{\lambda}_m^3} \left( |\alpha_m| + |\beta_m| \right) \leq \left( \sum_{m=1}^{\infty} \frac{1}{\tilde{\lambda}_m^6} \right)^{1/2} \left( \sum_{m=1}^{\infty} \left( |\alpha_m| + |\beta_m| \right)^2 \right)^{1/2} < \infty.$$

*B.4. Conclusion.* The absolute and uniform convergence of the majorizing series implies that the series for $u_{\text{app}}$ and all its termwise derivatives converge absolutely and uniformly on $[r_1, r_2] \times [0, T]$ for any finite $T$. Hence termwise differentiation is justified, and the limit function satisfies the wave equation (2.1), the Dirichlet boundary conditions (2.2), and the initial conditions (2.3) (by construction of the coefficients). This completes the proof.

## Appendix C. Oscillatory Integral Estimates and Asymptotic Orthogonality

In this appendix we collect several auxiliary estimates used in Section 7 in order to justify the asymptotic reduction of the exact spectral interactions to their semiclassical Liouville–Green form. The purpose of the appendix is not to develop an independent asymptotic theory but rather to provide a compact technical justification for the decay of off-diagonal spectral interactions arising in the semiclassical regime. The analysis follows the standard asymptotic Liouville–Green framework and nonstationary phase methodology developed in [6,7].

Throughout the appendix, we assume that the phase functions are sufficiently smooth on the finite interval $[r_1, r_2]$ and satisfy the nondegeneracy condition

$$\left|\Theta'(r)\right| \geq c_0 > 0.$$

*C.1. A nonstationary phase lemma.* We begin with a standard oscillatory estimate.

**Lemma C.1.** *Let*

$$I(\lambda) = \int_{r_1}^{r_2} a(r) e^{i\lambda\Theta(r)} dr,$$

*where*

$$a \in C^1[r_1, r_2], \qquad \Theta \in C^2[r_1, r_2],$$

*and suppose that* $\left|\Theta'(r)\right| \geq c_0 > 0$ *on* $[r_1, r_2]$. *Then*

$$I(\lambda) = O(\lambda^{-1}), \qquad \lambda \to \infty.$$

*Proof.* Integrating by parts, we obtain

$$I(\lambda) = \left.\frac{a(r) e^{i\lambda\Theta(r)}}{i\lambda\Theta'(r)}\right|_{r_1}^{r_2} - \frac{1}{i\lambda}\int_{r_1}^{r_2} \frac{d}{dr}\left(\frac{a(r)}{\Theta'(r)}\right) e^{i\lambda\Theta(r)} dr.$$

Since $|\Theta'(r)| \geq c_0 > 0$, all coefficients remain bounded on the finite interval $[r_1, r_2]$. Therefore,

$$|I(\lambda)| \leq \frac{C}{\lambda},$$

which proves the statement.

*C.2. Oscillatory mixed spectral contributions.* We now apply Lemma C.1 to the mixed semiclassical spectral terms arising in the Liouville–Green approximation.

Let

$$R_m^{(LG)}(r) = A_m(r)e^{i\Theta_m(r)} + \overline{A_m(r)}e^{-i\Theta_m(r)}$$

denote the corresponding Liouville–Green approximation in the high-mode regime introduced in Section 6.

For $m \neq n$, the mixed scalar products contain rapidly oscillating phase factors appearing in semiclassical spectral asymptotics [9]

$$J_{mn} = \int_{r_1}^{r_2} R_m^{(LG)}(r) R_n^{(LG)}(r) \rho(r) dr.$$

After decomposition into oscillatory components, the mixed terms reduce to integrals of the type

$$\int_{r_1}^{r_2} A(r) e^{i(\Theta_m(r) - \Theta_n(r))} dr,$$

where $A(r)$ is a smooth amplitude function.

For sufficiently large modes, the corresponding phase difference satisfies

$$|\Theta'_m(r) - \Theta'_n(r)| \geq c_{mn} > 0, \qquad m \neq n,$$

uniformly on $[r_1, r_2]$. Therefore, Lemma C.1 implies the estimate

$$J_{mn} = O\left(|m-n|^{-1}\right), \qquad |m-n| \to \infty.$$

Consequently, the off-diagonal spectral interactions vanish asymptotically in the high-mode limit.

*C.3. Relation to the exact spectral problem.* We finally explain the role of the oscillatory estimates in the exact spectral representation.

According to the Liouville–Green asymptotics established in Section 6, the exact

eigenfunctions admit the representation

$$R_m(r) = R_m^{(LG)}(r) + \varepsilon_m(r), \quad \varepsilon_m(r) \to 0, \quad m \to \infty,$$

in the corresponding weighted norm.

Substituting this decomposition into the exact orthogonality relation

$$\int_{r_1}^{r_2} R_m(r) R_n(r) \rho(r) dr = 0, \qquad m \neq n,$$

one obtains the decomposition

$$\int_{r_1}^{r_2} R_m^{(LG)}(r) R_n^{(LG)}(r) \rho(r) dr + R_{mn},$$

where the remainder term $R_{mn}$ contains the asymptotically small correction terms generated by $\varepsilon_m(r)$.

The oscillatory integrals appearing in the leading semiclassical contribution are controlled by the estimates above, while the remainder terms vanish due to the convergence

$$\varepsilon_m(r) \to 0.$$

Therefore, the semiclassical Liouville–Green orthogonality structure asymptotically reproduces the exact orthogonality relations of the Sturm–Liouville eigenfunctions.

This justifies the asymptotic reduction used in Section 7 and demonstrates that the semiclassical orthogonality structure emerges asymptotically from the exact spectral structure of the radial operator.